\documentclass[aps,prb,twocolumn,floatfix]{revtex4} 

\usepackage{graphics}

\begin{document} 

\title{Matrix Element and Strong Electron Correlation Effects in ARPES
from Cuprates}

\author{ A. Bansil$^1$, R.S. Markiewicz$^1$, C. Kusko$^1$, M.
  Lindroos$^{1,2}$, and S. Sahrakorpi$^1$}

\affiliation{ 1. Physics Department, Northeastern University, Boston
  Massachusetts 02115, USA
  \\
  2. Institute of Physics, Tampere University of Technology, 33101
  Tampere, Finland}


\begin{abstract}
  
  We discuss selected results from our recent work concerning the
  ARPES (angle-resolved photoemission) spectra from the cuprates. Our
  focus is on developing an understanding of the effects of the ARPES
  matrix element and those of strong electron correlations in
  analyzing photointensities. With simulations on
  Bi$_2$Sr$_2$CaCu$_2$O$_{8+\delta}$ (Bi2212), we show that the ARPES
  matrix element possesses remarkable selectivity properties, such
  that by tuning the photon energy and polarization, emission from the
  bonding or the antibonding states can be enhanced. Moreover, at low
  photon energies (below 25 eV), the Fermi surface (FS) emission is
  dominated by transitions from just the O-atoms in the CuO$_2$
  planes. In connection with strong correlation effects, we consider
  the evolution with doping of the FS of
  Nd$_{2-x}$Ce$_x$CuO$_{4\pm\delta}$ (NCCO) in terms of the
  $t$-$t'$-$U$ Hubbard model Hamiltonian. We thus delineate how the FS
  evolves on electron doping from the insulating state in NCCO. The
  Mott pseudogap is found to collapse around optimal doping suggesting
  the existence of an associated quantum critical point.

\end{abstract}


\maketitle

\section{Introduction}

Angle resolved photoemission spectroscopy (ARPES) has proven to be a
powerful technique for probing the properties of the superconducting
cuprates. In analyzing the data, it is important to keep in mind the
effects of the ARPES matrix element as well as those of the strong
electron correlations\cite{bansil99,lindroos02,bansil02,asensio03,%
sahrakorpi03,chuang03,kordyuk02,KMLB,MK4,damascelli03}. With this
motivation, we present in this article some of our recent work
directed at understanding these effects in the cuprates.

In connection with matrix element effects, we discuss the ARPES
intensity from initial states in the vicinity of the $M(\pi,0)$
symmetry point in tetragonal Bi2212. For this purpose, one-step
photointensity calculations for several different initial state
energies for both the bonding and the antibonding bands over the
photon energy range of 10-100 eV are presented. These results give
insight into how the cross-sections for exciting these bands vary with
the photon energy on the one hand, and the character and energy of the
initial state on the other hand. We comment on systematics in the
theoretical spectra, and provide examples of specific energies or
energy ranges, which are well-suited for highlighting various
interesting spectral features. We also consider the extent to which
different sites in the unit cell contribute to the photointensity.
This question is addressed via computations of the dipole matrix
element for exciting the bonding or the antibonding portion of the
Fermi surface of Bi2212 throughout the Brillouin zone for two
different polarizations of the incident light. Here the results show a
remarkable selectivity of the spectra with respect to the O-sites for
low photon energies. On the whole, our computations indicate that by
fine tuning photon energy and/or polarization, ARPES can help zoom in
on specific states and/or sites in the cuprates. We note that the
methodology used here has been described in detail in our earlier
publications on the
cuprates\cite{bansil95,lindroos95,bansil99,lindroos02,bansil02,%
asensio03,sahrakorpi03,chuang03}. The crystal potential used in Bi2212
is the same as that used in Ref.~\onlinecite{lindroos02} and,
consistent with ARPES data, it does not contain Bi-O pockets around
the $M$-point.

Turning to the electron doped cuprates, we discuss the properties of
the $t$-$t'$-$U$ Hubbard model Hamiltonian, keeping in mind the recent
doping dependent ARPES data on NCCO. The analysis is carried out using
the framework of the mean-field Hartree-Fock (HF) as well as that of
the self-consistent renormalization (SCR) theory to include the effect
of fluctuations. The relevant details of methodology are given in
Refs.~\onlinecite{KMLB},~\onlinecite{MK4}, and~\onlinecite{Mor}.  It
should be noted that our comparisons between theory and experiment on
NCCO do not include the effects of the ARPES matrix element, but are
considered to be adequate for our purposes of gaining a handle on the
overall topology of the FS. An effort to include the ARPES matrix
element in the presence of strong correlations is in progress and will
be taken up elsewhere. The experimental band dispersions are used to
fit the doping dependence of $U$, which is found to be quite similar
for the HF and the SCR computations. In this way, we determine the
evolution of the FS and the Mott pseudogap in NCCO. This study also
gives insight into the stability of the uniformly doped phase (with
respect to the formation of competing nano-scale orders) upon electron
vs. hole doping.

\section{Selectivity Properties of the ARPES Matrix Element in Bi2212}

\begin{figure}
\begin{center}  
    \resizebox{8.5cm}{!}{\includegraphics{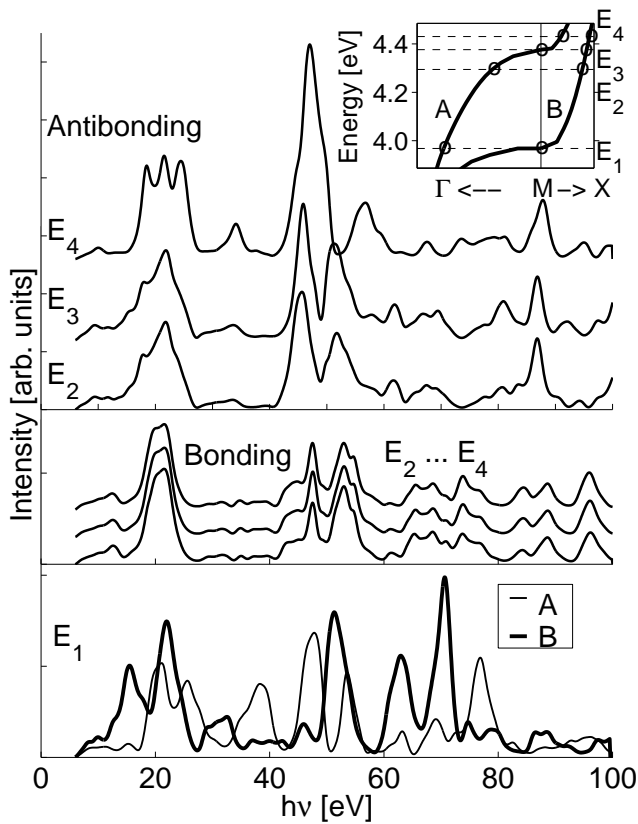}} 
\end{center}
\caption{ Theoretical one-step ARPES photointensities for emission
  from the antibonding (A) and bonding (B) bands in the vicinity of
  the $M(\pi,0)$-point as a function of the photon energy $h\nu$ in
  Bi2212. Insert shows the A and B bands and four initial state
  energies $E_1$-$E_4$ at which results for intensities are shown
  (open circles mark the relevant ${\mathbf k}_\parallel$-values).
  Incident light is assumed to be polarized along the $x$-axis,
  defined as the direction of the Cu-O bonds. }
\label{fig:1}
\end{figure}
Figure~\ref{fig:1} presents one-step ARPES photointensities for
tetragonal Bi2212 in the vicinity of the $M(\pi,0)$-symmetry point at
four illustrative initial state energies. The ${\mathbf
k}_\parallel$-values are chosen for each energy so that the
antibonding (A) or the bonding (B) band is excited as indicated by
open circles in the insert. At energies $E_1$ and $E_2$, the A-band is
electron-like around $\Gamma$, while at $E_4$, it is hole-like around
$X(\pi,\pi)$. Referring to energies, $E_2$ through $E_4$, the six
topmost curves in the figure show that the ARPES intensity from either
the A or the B band is more or less similar for small (compared to the
bilayer splitting) changes in energy. However, the intensity is seen
to vary dramatically with photon energy and these variations extend up
to the 100 eV upper energy limit considered. For example, the A-band
is quite intense in energy ranges around 20 eV, 47 eV and 88 eV. The
relative intensities of the A and B bands are seen to differ greatly.
For example, A is far more intense than B at around 47 eV and 88 eV,
while the case is opposite around 95 eV. Such special photon energies
can help focus on the properties of the A or the B band in the ARPES
spectra. The theoretically predicted enhancement of the A to B
features at 47 eV has been exploited recently by Chuang et
al.\cite{chuang03} to resolve bilayer splitting throughout the doping
range from overdoped to underdoped samples, and to adduce that some
coherence of electronic states between the CuO$_2$ planes persists in
Bi2212 even in the underdoped regime.

The preceding discussion makes it clear that the ARPES cross-sections
for excitation of states near the Fermi energy ($E_F$) depend strongly
on photon energy and that these photon energy dependencies differ
substantially between the A and B type states. It is interesting as
well to consider how these cross-sections vary with the initial state
energy. A sensible energy scale in this connection is that provided by
the bilayer splitting, $\Delta_{\text{bilayer}}$. The superconducting
energy scale, $\Delta_{\text{super}}$, is $\approx$ 30-50\% of
$\Delta_{\text{bilayer}}$. For this purpose, we compare results in
Figure~1 at the energy $E_1$ (lowest panel) with those at $E_2-E_4$
(upper panels), where $E_1$ is separated from $E_2-E_4$ by about
$\Delta_{\text{bilayer}}$. We see, for example, that in going from
$E_4$ to $E_1$, the A-band develops peaks at around 38 and 77 eV,
while the B-band displays new peaks at around 18, 63 and 70 eV. On the
other hand, some energy ranges are less sensitive to changes in the
initial state energy, e.g. the region from 60-75 eV for the A-band and
the regions from 35-45 eV and above 75 eV for the B-band. These
considerations will be important in identifying energy regions, which
may be particularly suitable for implementing recent
proposals\cite{vekhter02} for developing ARPES as a self-energy
spectroscopy, where the role of ${\mathbf k}_\parallel$ and energy
dependencies of the ARPES matrix element needs to be minimized.

\begin{figure}
\begin{center}  
    \resizebox{8.5cm}{!}{\includegraphics{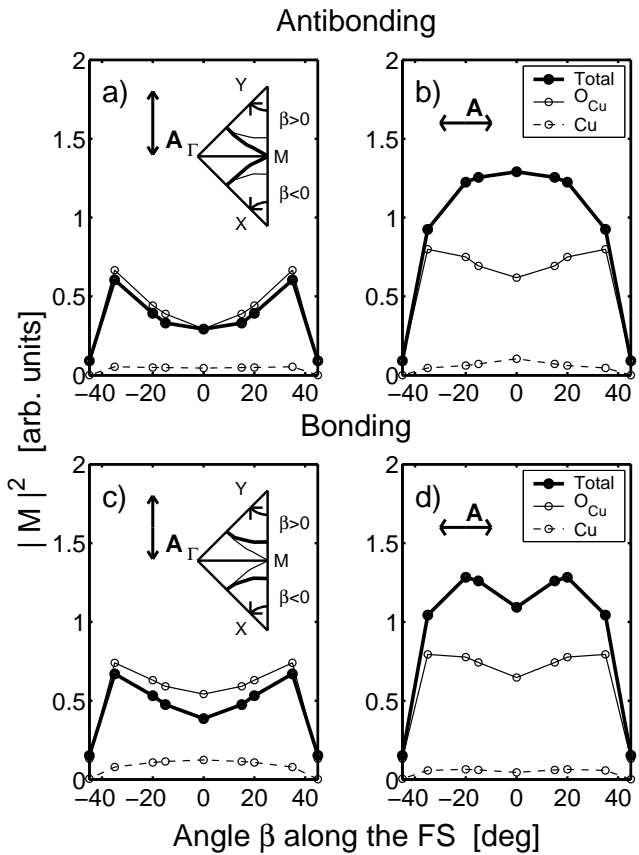}} 
\end{center}
\caption{ ARPES intensity defined as the square of the dipole matrix
  element, ${|M|}^2$ (see Eq.~\ref{eq:1}), for excitation of initial
  states lying along the FS of Bi2212. Thick solid lines give total
  intensity, thin solid lines give partial contribution from just the
  O$_{Cu}$ sites in the CuO$_2$ bilayers; the corresponding partial
  contribution from Cu sites is shown dashed. Photon energy is 22 eV.
  The right hand panels refer to light polarized along the Cu-O bond
  in $x$-direction (denoted by the double arrow of the vector
  potential ${\mathbf A}$), while in the left hand panels the
  polarization vector is rotated by 90$^o$. (a) and (b) consider
  excitation of states on the antibonding FS [thick solid line in the
  insert to (a)]. The ${\mathbf k}_\parallel$ values along the FS are
  defined by the angle $\beta$ (inserts in a,c), where $\beta =0^o$
  corresponds to the $X(Y)-M$-direction, and $\beta =\pm 45^o$
  corresponds to the $Y(X)-\Gamma$ direction. (c) and (d) similarly
  refer to the bonding FS. All intensities are normalized correctly
  relative to one another.}
\label{fig:2}
\end{figure}
The ARPES matrix element also exhibits remarkable site-selectivity
properties in Bi2212 in that, for low photon energies ($h\nu \leq 25$
eV), the intensity for emission around $E_F$ is dominated by
excitations from just the O-sites in the CuO$_2$ planes, even though
the relevant initial states possess an admixture of Cu and O
character. Our analysis suggests that at higher photon energies, the
contribution to intensity from Cu sites increases, and at around 40
eV, the Cu and O contributions are roughly
comparable.\cite{sahrakorpi03} As discussed
elsewhere\cite{lindroos02,sahrakorpi02,sahrakorpi03}, insight in this
regard can be obtained by considering the square of the dipole matrix
element, i.e.
\begin{equation}
{ | M | }^2 \equiv {| < \Psi_f | {\mathbf A} \cdot {\mathbf p} |
\Psi_i > |}^2
\label{eq:1}
\end{equation}
where ${\mathbf A}$ is the vector potential of the incident photon
field, ${\mathbf p}$ is the momentum operator, and $\Psi_i$ and
$\Psi_f$ are the bulk crystal wavefunctions of the initial and final
states involved.  Fig.~\ref{fig:2} presents the intensity based on
Eq.~\ref{eq:1} for emission from the antibonding as well as the
bonding portions of the FS for two different polarizations of light.
The contributions to the total intensity arising from the O-sites
(denoted by O$_{Cu}$/thin solid line) and the Cu sites (thin dashed
line) in the CuO$_2$ bilayers are shown. The contributions from other
sites (not shown) are not always negligible as seen by comparing the
total intensity curves (thick solid lines) with the sum of the thin
curves, a point to which we return below. Incidentally, O contribution
in Figs.~\ref{fig:2}(a) and~\ref{fig:2}(c) is slightly higher than the
total intensity curve -- this reflects interference effects between
the contributions from different sites when the absolute value of the
matrix element is taken in Eq.~\ref{eq:1}.

In looking at the results of Fig.~\ref{fig:2}, it is important to bear
in mind the effect of symmetry on the dipole matrix element. In view
of the direction of the polarization vector, the left hand side panels
pick up only the part $M_y \equiv <\Psi_f|p_y|\Psi_i>$ of the momentum
operator, ${\mathbf p}$, while the right hand side panels similarly
yield $M_x \equiv <\Psi_f|p_x|\Psi_i>$.  Comparing the total
intensities in (a) with (c), or in (b) with (d), we see thus that the
squared absolute value of $M_y$ is smaller than that of $M_x$ for the
antibonding as well as the bonding FS in the vicinity of the
$M(\pi,0)$-point. It is straightforward to show that the symmetry of
the curves in Fig.~\ref{fig:2} around $\beta=0$ is tied to the mirror
symmetry of the underlying tetragonal lattice through the $\Gamma-M$
line. The mirror symmetry through the $\Gamma-Y$ and $\Gamma-X$ lines,
on the other hand, can be used to deduce that at $\beta = \pm 45^o$,
$|M_x|=|M_y|$, so that the intensity at $\beta = \pm 45^o$ is the same
for either horizontal or vertical polarization, as is seen to be the
case with reference to various panels of Fig.~\ref{fig:2}.

Fig.~\ref{fig:2} shows that the Cu contribution to the total intensity
is quite small at $\sim 22$ eV photon energy for either polarization
for both the bonding and the antibonding FS states throughout the
Brillouin zone. The O$_{Cu}$ contribution is seen to be substantial in
all four cases. This is particularly the case for polarization along
the $y$-direction in panels (a) and (c), suggesting that in the
vicinity of the $M(\pi,0)$-point the $A_y$ polarized light would be
well-suited for probing the character of O$_{Cu}$ electrons.  The
results for the $A_x$-polarized light, in the right hand side panels
(b) and (d) show a somewhat more complicated behavior in that
significant contribution from other atoms (than O$_{Cu}$) in the unit
cell is involved. Our analysis indicates that these additional
contributions arise from O atoms in the Bi-O and Sr-O layers in the
structure. Despite some uncertainty with respect to the extent to
which Bi contributes to emission from the $E_F$ in Bi2212, it seems
then that $A_x$ polarization will likely be more sensitive for
investigating the properties of apical O-sites in the Sr-O planes.

\section{Mott Gap Collapse in NCCO}

\begin{figure}
\begin{center}  
    \resizebox{8.5cm}{!}{\includegraphics{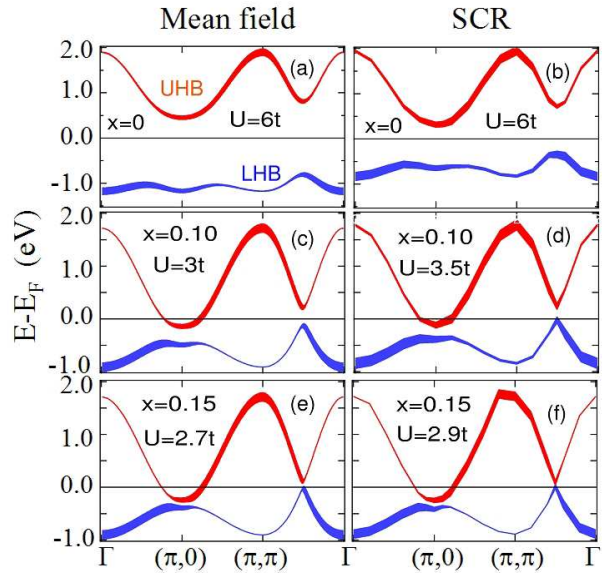}} 
\end{center}
\caption{Dispersion curves for NCCO within the $t$-$t'$-$U$ Hubbard
  model, showing the evolution of the upper and lower Hubbard bands
  (denoted by UHB and LHB) with doping $x$. Results for the mean field
  computations are given in the left column, and for the mode
  coupling, self-consistent renormalization (SCR) theory in the right
  column. Values of $x$ and $U$ are marked in each panel. Thickness of
  bands denotes their spectral weight. Horizontal lines mark the Fermi
  energy.}
\label{fig:3}
\end{figure}
We now discuss briefly the issue of evolution of the electronic
structure and FS of NCCO with doping and the related collapse of the
Mott gap around optimal doping with reference to Figs.~3 and~4. As
already noted, these results are based on the $t$-$t'$-$U$ Hubbard
model Hamiltonian, where the value of $U$ has been used as a fitting
parameter to reproduce the experimentally observed doping dependence
of the FS in NCCO observed via ARPES experiments.

The mean field\cite{KMLB} and the SCR\cite{MK4} results of
Fig.~\ref{fig:3} are seen to be quite similar, although the specific
values of $U$ needed in the two cases are somewhat different.  The SCR
computations are of course more satisfactory since the effects of
fluctuations are incorporated.  The mean field theory suffers from the
well-known problem that it predicts the Neel temperature to be too
high. However, when fluctuations are included (in the mode coupling
approach), true long-range order only appears at $T=0$, consistent
with the Mermin-Wagner theorem, while a \emph{pseudogap} opens up
close to the mean-field Neel transition due to short-range
antiferromagnetic order.  The undoped system ($x=0$) displays a wide
gap between the UHB and the LHB. At intermediate doping ($x=0.10$),
the band gap is substantially reduced and electron pockets appear
around the $(\pi,0)$-point in the UHB. At optimal doping ($x=0.15$),
the band gap has essentially disappeared, and recalling that thin
lines possess little spectral weight in Fig.~3, the band dispersions
resemble those for the uncorrelated case with a large hole FS sheet
centered around the $(\pi,\pi)$-point. It is important to observe that
our analysis indicates that the Hubbard $U$ must \emph{decrease}
significantly with doping -- by about a factor of two from the value
of $6t$ in the undoped case to around $3t$ in the optimally doped
system\cite{Kana}.  Interestingly, our finding that the Mott pseudogap
collapses around optimal doping, where the staggered magnetization
goes to zero and the $T=0$ Neel ordered state terminates, suggests the
presence of an associated quantum critical point (QCP) in the
spectrum. The manifestation of optimal superconductivity in the
vicinity of this QCP is consistent with results in other magnetic
materials\cite{MaLo}, including hole-doped cuprates\cite{Tal}.

\begin{figure}
\begin{center}  
    \resizebox{7.5cm}{!}{\includegraphics{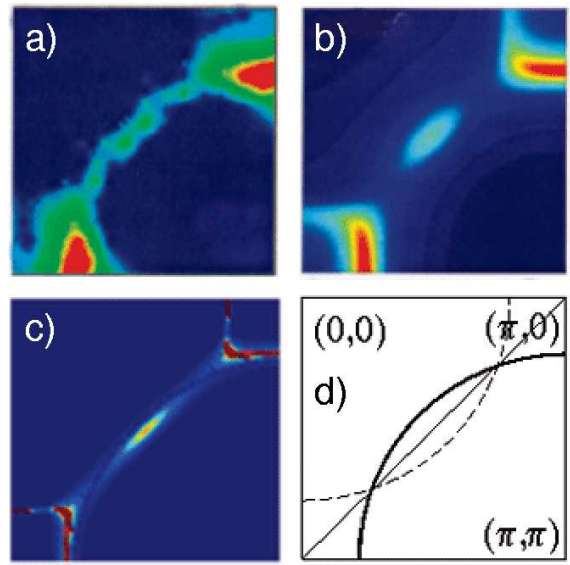}} 
\end{center}
\caption{Fermi surface of NCCO at $x=0.10$. Experimental ARPES results
  for emission from the Fermi energy in (a) are compared with mean
  field (b), and SCR (c) computations. Frame (d) shows a schematic
  uncorrelated FS (thick solid line) and the related "shadow" FS
  (dashed line) obtained by folding the solid line around the magnetic
  zone boundary. }
\label{fig:4}
\end{figure}
Insight into the preceding discussion is provided by the illustrative
example of Figure~4, which compares computed FS maps for the mean
field (top right) and SCR (bottom left) calculations, with the
corresponding experimental\cite{nparm} ARPES intensities (top left) at
intermediate doping $x=0.10$. In the mean-field calculations, a small
second-neighbor hopping $t''$ has been included to better fit the
shape of the FS near the nodal point, and the data are broadened to
reflect experimental resolution. We emphasize that the effects of the
ARPES matrix element, discussed in the preceding section, are not
included in these computations. Nevertheless, we expect accord between
theory and experiment at the level of providing the correct topology
of the FS. This indeed is seen to be the case. Recalling Fig.~3, both
theory and experiment display the electron pocket around $(\pi,0)$
[with a symmetrically placed pocket located at $(0,\pi)$], arising
from the doping of the UHB.  Additionally, as the Mott pseudogap
decreases, a second \emph{hole}-like Fermi surface appears near
$(\pi/2,\pi/2)$ coming from the LHB. Notably, evidence for two-band
conduction has been found in this doping range.\cite{jiang94}

\section{Summary and Conclusions}

In this article, we have briefly discussed two aspects of our recent
work aimed at understanding the high-$T_c$ cuprates via ARPES
experiments: (i) Matrix element effects in ARPES spectra of Bi2212,
and (ii) Evolution of electronic structure and FS of NCCO with
electron doping.

In considering matrix element effects, we discuss two distinct types
of selectivity properties of the ARPES matrix element. Firstly,
depending on the character of the initial state involved, the ARPES
matrix element can yield widely differing cross-sections as a function
of the energy and polarization of the incident photons. This effect
can be used to discriminate, for example, between the bonding and
antibonding pieces of the FS in Bi2212. Such a theoretically predicted
enhancement of antibonding to bonding intensity at 47 eV has in fact
been exploited in Ref.~\onlinecite{chuang03} to adduce that the
bilayer splitting in Bi2212 is surprisingly insensitive to doping and
that some coherence of electronic states persists across the CuO$_2$
bilayers even in the underdoped regime.  Secondly, we show that the
ARPES matrix element possesses remarkable "site-selectivity" in that
the emission in the low photon energy range (5-25 eV) in Bi2212 is
dominated by excitations from just the O-sites in CuO$_2$ planes
throughout the $(k_x,k_y)$-plane. The polarization vector of the
incident light may provide some sensitivity to O-sites in Bi-O and
Sr-O planes. At higher photon energies (above 25 eV), Cu sites begin
to progressively contribute to the ARPES intensity. These results
suggest that ARPES could potentially allow one to focus on the
properties of specific groups of electrons associated with various
sites in the unit cell.

With regard to the electron doped cuprates, we have analyzed the
evolution of the electronic structure and the FS of NCCO with doping
within the framework of the $t$-$t'$-$U$ Hubbard model Hamiltonian in
the light of related ARPES experiments. Computations have been carried
out both within the mean field Hartree Fock as well as the
self-consistent renormalization frameworks. We show how the FS
develops when electrons are added in the undoped insulating state: At
first the electrons enter the upper Hubbard band around the $(0,\pi)$
and $(\pi,0)$ points, but on further doping electrons also enter the
lower Hubbard band first around the $(\pi/2,\pi/2)$ point. Finally,
around optimal doping, the FS essentially resembles that of the
uncorrelated state with a large $(\pi,\pi)$ centered hole sheet.
Notably, we find that the effective Hubbard $U$ must decrease
substantially with doping, such that the Mott pseudogap collapses
around optimal doping, suggesting the existence of an associated
quantum critical point in the spectrum. Our study also indicates that
a uniformly doped antiferromagnetic state in the cuprates is likely
more easily accessible via electron doping rather than through
hole-doping, since in the latter case the system is unstable towards
various competing orders (nano-scale phase separations).  \\

\begin{acknowledgments}
  
  This work is supported by the US Department of Energy contract
  DE-AC03-76SF00098, and benefited from the allocation of
  supercomputer time at NERSC, Northeastern University's Advanced
  Scientific Computation Center (ASCC), and the Institute of Advanced
  Computing (IAC), Tampere.  One of us (S.S.) acknowledges Suomen
  Akatemia and Vilho, Yrj\"o ja Kalle V\"ais\"al\"an Rahasto for
  financial support.

\end{acknowledgments}

\end{document}